%% file: main-lncs.tex
\documentclass[runningheads]{llncs}
\usepackage[T1]{fontenc}
\usepackage{babel}
\usepackage[babel]{microtype}
\usepackage[babel]{csquotes}
\usepackage{siunitx} 

\usepackage[svgnames]{xcolor}
\usepackage{graphicx}

\usepackage{amsmath}
\usepackage{amssymb}
\usepackage{amsfonts}
\usepackage{mathtools}
\usepackage{dsfont}

\usepackage{hyperref}
\usepackage[capitalize]{cleveref}

\usepackage{newtxtext}       %
\usepackage[varvw]{newtxmath}       

\usepackage{graphicx}
\input{setup-notation.tex}

\begin{document}
\title{Identification Codes for Ad Hoc Networks: Use Cases and Challenges}
%
%
\author{Athin Mohan\inst{1}\orcidID{0009-0001-3651-1377} \and
Roberto Ferrara\inst{1}\orcidID{0000-0002-1991-3286} \and
Christian Deppe\inst{1}\orcidID{0000-0002-2265-4887}}
\authorrunning{A. Mohan et al.}
%
\institute{Institute for Communications Technology, Technische Universität Braunschweig, Germany
\email{\{athin.mohan, roberto.ferrara, christian.deppe\}@tu-braunschweig.de}}
\maketitle              
\begin{abstract}
In this study, we focus on large-scale ad hoc networks characterized by a high node density. To achieve energy efficiency, the devices must minimize their data transmission, thereby reducing the total number of bits sent and the associated energy consumption. We consider using ID codes for the address field in the MAC layer and reduce the header size. The work explores a range of use cases and challenges that demonstrate the benefits of incorporating these ID codes into packet routing mechanisms for ad hoc networks, thereby providing a foundation for understanding their applicability and the necessary design adaptations.

\end{abstract}
\input{sec_introduction.tex}

\input{sec_prerequisites.tex}

\input{sec_systemmodel.tex}

\input{sec_challenges.tex}

\begin{credits}
\subsubsection{\ackname} This work was primarily carried out within the planned research activities of the QDCamnetz project, funded by the Federal Ministry of Research, Technology and Space (BMFTR) of Germany under grant number 16KISQ0170~(CD). It also benefited from research results and scientific discussions arising from other BMFTR-funded projects: the joint project \emph{6G-life$^2$} within the ``Souverän. Digital. Vernetzt.'' program under grant number 16KIS2415~(AM, CD), as well as projects funded under grant numbers 16KISQ169~(AM, CD), 16KIS2602~(AM, RF, CD), 16KISQ028~(CD), 16KISQ038~(CD), 16KISR038~(CD), and 16KIS2234~(CD). The authors gratefully acknowledge this support.
\end{credits}

\bibliographystyle{splncs04}
\bibliography{bibliography}
\end{document}

%% file: setup-notation.tex
\newcommand*{\numofnodes}{\ensuremath{N}}

%% file: sec_introduction.tex
\section{Introduction}

Identification Codes (ID) was introduced by Ahlswede and Dueck~\cite{ahlswede1989identification}. ID offers a fundamentally different scaling behavior compared to Shannon's theory of transmission. While Shannon’s theory shows that the number of transmittable messages grows exponentially with block length and rate, identification allows the number of identifiable identities to scale double-exponentially under the same conditions. Consequently, ID represents a specialized framework within the broader Beyond-Shannon communication paradigm. Recent works on ID have outlined its application in various domains including wake-up signals~\cite{perotti2024identification}, molecular communication~\cite{zhao2025identification} and nanorobot swarms~\cite{labidi2026joint}. 
While identification theory is primarily developed in an asymptotic setting, practical implementations have already been explored. Researchers have proposed various real-world constructions using structured codes like Reed-Solomon and Reed-Muller, combinatorial designs, and pseudo-random methods that show identification can work effectively even with finite blocklengths, limited hardware, and strict latency requirements. These approaches suggest that ID isn’t just a theoretical curiosity, but a viable option for real systems~\cite{ferrara2022implementation,von2025codes,perotti2024identification}.

ID coding shares strong similarities with common functions like checksums, cyclic redundancy checks (CRCs), and non-cryptographic hash functions~\cite{von2025codes}. What ties them together is two key features: first, they compress a message into a shorter codeword, and second, they rely on a verification step where this codeword is compared against a received or expected version. Notably, all these functions are inherently lossy. They discard information in such a way that multiple different messages can produce the same codeword, making exact reconstruction impossible. This property aligns closely with the core idea behind identification, where the goal isn’t to recover the original message, but to determine whether a specific one is present.

Ad hoc networks describe networks that are composed of mobile devices or nodes that spontaneously establish temporary connections on the fly, without relying on any pre-existing infrastructure or centralized management. Wireless ad hoc networks, by their decentralized design, are well-suited for scenarios where centralized control is impractical or unavailable. This architecture can enhance scalability compared to traditional managed wireless networks. Their ability to operate with minimal setup and be deployed rapidly makes them ideal for emergency response situations such as natural disasters or military operations, where infrastructure may be damaged or absent. The use of dynamic, self-adapting routing protocols further supports fast network formation and resilience in changing environments.

In this work we explain how ID codes can be useful in reducing the header size in ad hoc networks. This is especially beneficial for energy limited ad hoc networks such as wireless sensor networks~\cite{akyildiz2002wireless}. We provide concrete examples of use cases for the usage of ID in such networks. We find that a substantial reduction in header size is possible with the use of ID. Since common randomness is essential for the proposed scheme, we also discuss how common randomness can be distributed between pairs of senders and receivers.

We introduce the required preliminaries in~\cref{sec:prereq} about ID codes and the distinction between randomized and deterministic ID. \Cref{sec:packetrouting} give some preliminary examples for usage of ID codes for packet routing. We then detail how ID codes can be used for ad hoc networks in~\cref{sec:adhoc}. The challenges that are faced in implementing the solution and directions for future work are provided in~\cref{sec:concl}.

%% file: sec_prerequisites.tex
\section{Prerequisites}\label{sec:prereq}

\subsection{ID Code}

An $(n,N,\lambda_1,\lambda_2)$ ID code~\cite{ahlswede1989identification} consists of $N$ probability distributions, $p_i$, each defined over the input alphabet $X^n$, and $N$ corresponding decoding subsets, $D_i$, within the output alphabet $Y^n$  such that it satisfies the following for a channel $W^n$, $x^n \in X^n$ and $y^n \in Y^n$.
\begin{align}
    \sum_{x^n \in X^n} p_i(x^n) W^n(D_i|x^n) \ge 1 - \lambda_1 \quad \forall i \in [1: N], \\
    \sum_{x^n \in X^n} p_j(x^n) W^n(D_i|x^n) \le \lambda_2 \quad \forall i, j \ i \neq j \label{eq:type2}
\end{align}

The upper type-I error probability bound, denoted as $\lambda_1$, is influenced by the level of distortion introduced by the communication channel $W^n$. This relationship indicates that the probability of incorrectly identifying a transmitted codeword increases with the distortion level of the channel.
The other type of error in~\autoref{eq:type2} is the Type-II error that involve false-positive outcomes at the decoder, where the received codeword $y^n$ leads the receiver to incorrectly conclude that the source's ID $j$ matches its own ID $i$, even though they are different.

\begin{figure}
    \centering
    \includegraphics[width=0.80\linewidth]{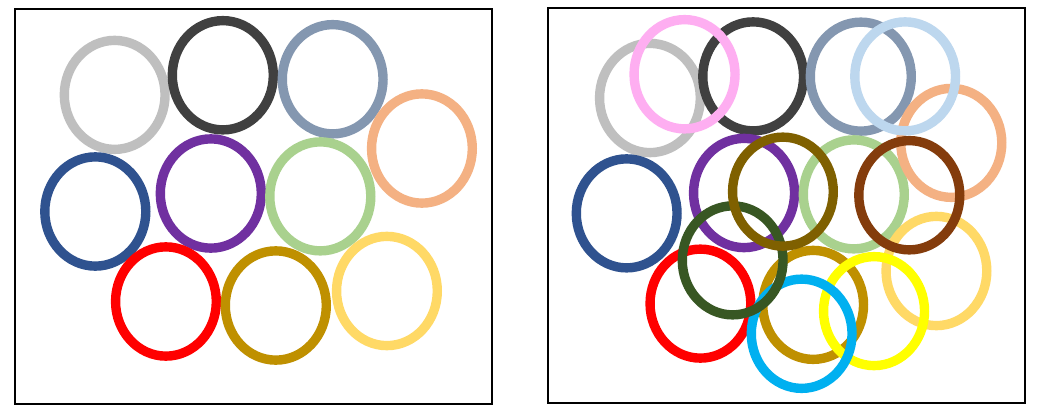}
    \caption{Decoding regions in Transmission (left) and Identification (right).\newline Each colour represents the decoding region of a codeword. The overlap between decoding regions in ID enables an increase in the number of identities for the same blocklength.}
    \label{fig:TxID_Decoding_Regions}
\end{figure}

The primary distinction between identification and transmission lies in the fact that the decoding regions $D_i$ do not need to be mutually exclusive as illustrated in \cref{fig:TxID_Decoding_Regions}. Consequently, this allows for a significantly greater number of messages to be supported. While message transmission scales exponentially with block length, message identification achieves a double-exponential scaling. This means that for a given block length $n$, the number of messages in transmission scales as $2^n$, whereas the number of IDs in identification scales as $2^{2^n}$. If $R$ denotes the ID rate,
\begin{equation}\label{eq:doubleexpo}
    N(n, \lambda_1, \lambda_2) = \exp\exp nR \ \Longleftrightarrow \ R = \frac{\log\log N}{n}
\end{equation}

\subsection{Randomized Identification vs Deterministic Identification}
The key difference between randomized~\cite{von2025codes} and deterministic identification codes~\cite{vorobyev2024deterministic} lies in the sender's ability to utilize local randomness during encoding. This difference significantly impacts the maximum code size and the rate at which the code can grow. Ahlswede and Dueck's groundbreaking work demonstrated that when using randomized encoding, the size of identification codes increases double-exponentially with the blocklength over discrete memoryless channels (DMCs), marking a significant shift from the single-exponential growth seen in traditional message transmission. Randomized encoding is crucial for achieving this double-exponential growth, whereas deterministic encoding results in an exponential scaling of identifiable messages over DMCs. Even though deterministic identification loses the double-exponential benefit, its rate of identification still surpasses the rate of message transmission on the exponential scale.

%% file: sec_systemmodel.tex
\section{Identification for Packet Routing}\label{sec:packetrouting}

\begin{figure}
    \centering
\includegraphics[width=0.80\linewidth]{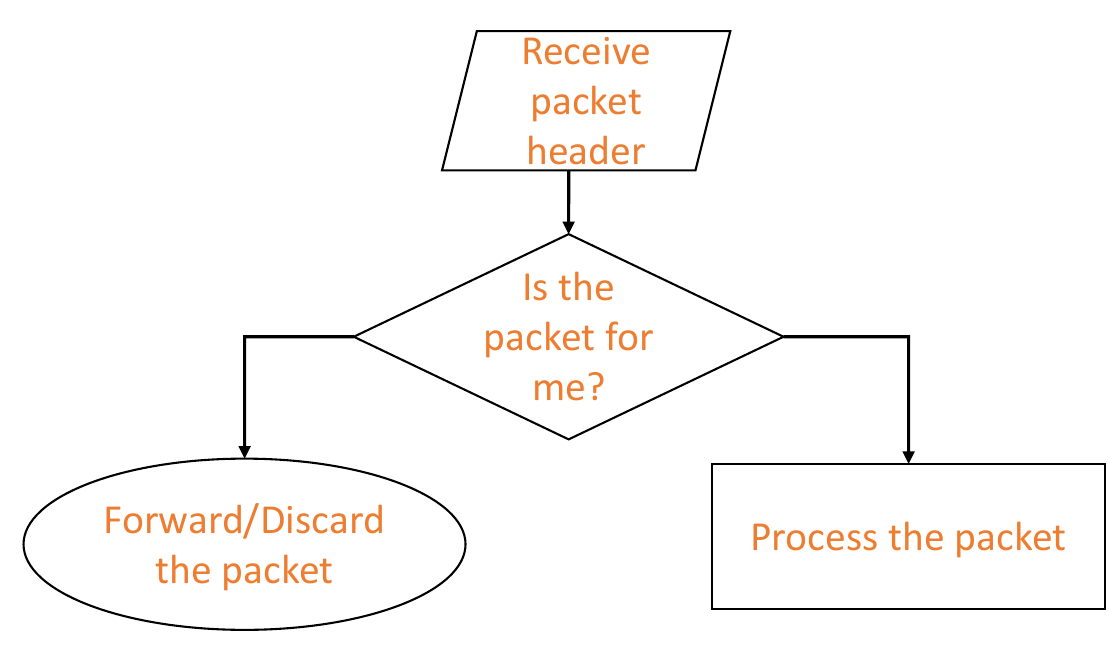}
    \caption{Identification Scenario in the packet routing example}
    \label{fig:flowchart}
\end{figure}

In packet routing within a network, MAC (Media Access Control) addresses are utilized in packets to facilitate communication at the data link layer (Layer 2) of the OSI (Open Systems Interconnection) model~\cite{kumar2014osi}. This ensures that data is correctly directed to the intended device within the same local network segment. We propose to use ID codes to identify the unique MAC addresses of the devices. The concept of encoded addresses not only allows for additional size reduction but also enhances the system's scalability. Essentially, the concept that we apply is that the receiver is only interested in knowing whether the data packet is for itself or not. In this case, it only wants an answer to the binary yes-or-no question, whether it is the intended recipient of the data packet. It is not interested in identifying the target node unless the target node is itself as is illustrated in \cref{fig:flowchart}.

\begin{figure}
    \centering
    \includegraphics[width=0.80\linewidth]{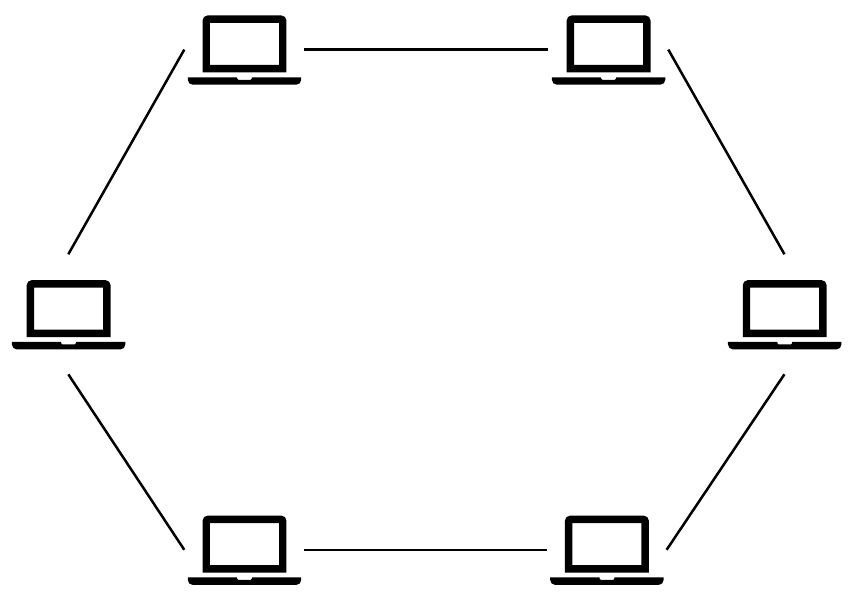}
    \caption{Ring topology of nodes}
    \label{fig:ring}
\end{figure}

A simple scenario where this can be easily seen to be useful is the ring topology as seen in~\cref{fig:ring}. In this scenario, the node which wants to send the data encodes the destination ID in the header and forwards the packet in the direction of the destination node. When an intermediate node receives the packet, it would decode the ID and upon knowing that the packet is not for it, forwards the packet in the opposite direction as it was received. This continues until the packet reaches the destination.

\begin{figure}
    \centering \includegraphics[width=0.80\textwidth]{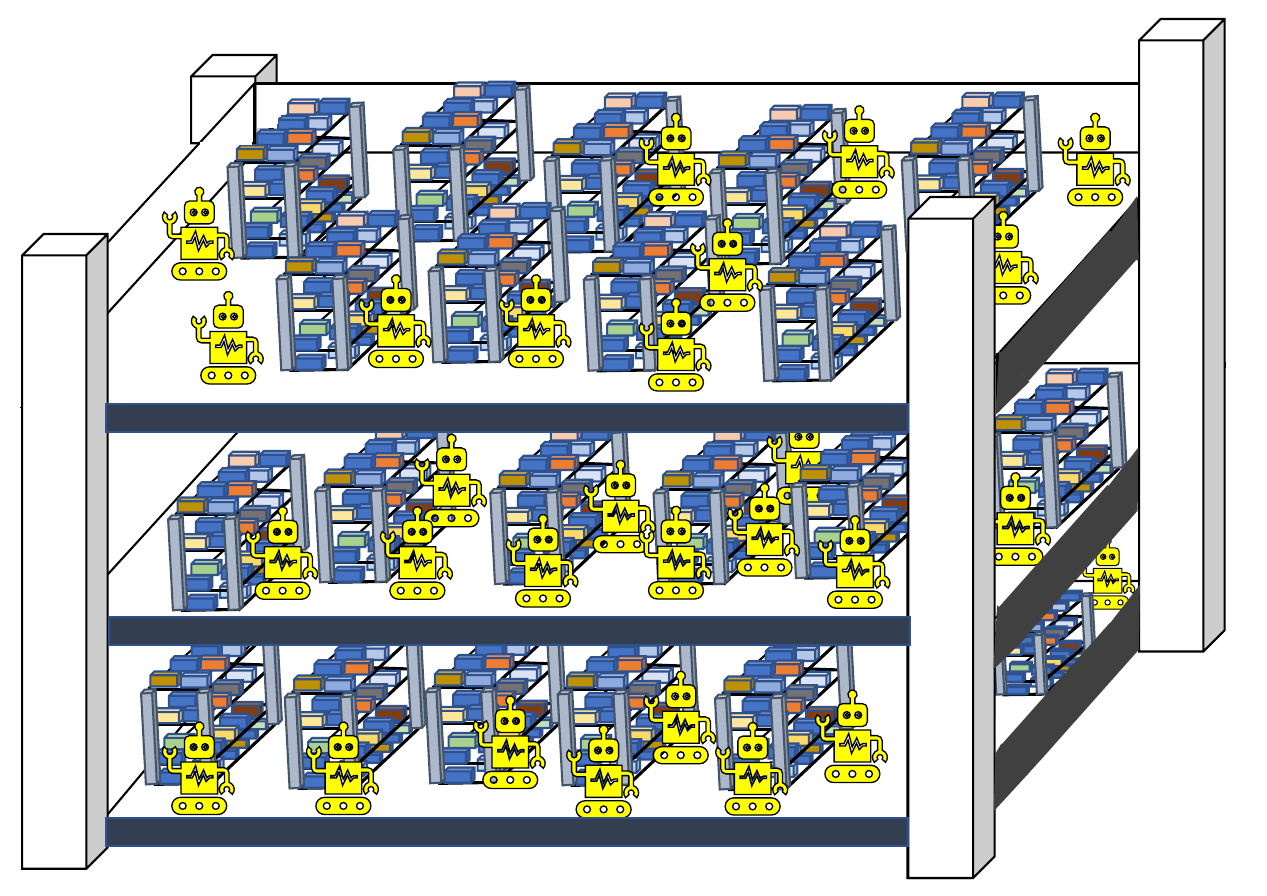}
    \caption{An example of a smart warehouse with mobile robots}
    \label{fig:example2}
\end{figure}

As a second example, let us consider a situation with a massive number of mobile nodes in a wireless network. An example could be a large-scale industrial environment such as smart warehouses with hundreds of autonomous mobile robots. Centralized routing becomes impractical due to dynamic topologies, high node mobility, and the prohibitive cost of maintaining global routing tables. Let each node contain a list of nodes it wants to communicate with and the nodes are distributed randomly such that the destination node might not be reachable within one hop. This means the destination is not reached when the source node broadcasts the packet. This is illustrated in \cref{fig:example2}. We provide a simple broadcasting-based solution to the problem based on Identification codes.   

When a node wants to send a message to any other node on its list, it would use the ID code of the MAC address of receiver and broadcast the packet. Any other node in its neighborhood that receives the packet checks whether the packet is for itself or not. If it is not intended for itself, it broadcasts the packet again. In this manner, repeated broadcasts allow the message to reach the intended receiver. This decentralized, self-checking approach ensures scalable and robust communication without requiring global knowledge, making it well-suited for real-time, infrastructure-less networks such as the example stated. The challenge for this flooding-based strategy is to tackle the problem of messages getting broadcast in a loop. This can be countered using Time-to-live or probabilistic forwarding.

\section{Identification Codes for Adhoc Networks}\label{sec:adhoc}
We consider an ad hoc network with \numofnodes\,nodes. Each node is mobile and connects and disconnects from the network independently. When a node typically joins the network and starts participating in the network, the following steps are performed.
\begin{itemize}
    \item Neighbour Discovery
    \item Updating Tables/Information for Routing
    \item Route Establishment
    \item Data forwarding
    \item Connectivity maintenance
    \end{itemize}

As is evident, neighbour discovery is an important step in the joining procedure as it allows the node to be discovered by other nodes and also detects nodes that it can connect to directly. During this phase, the node may perform the step by active probing or passive listening or a hybrid approach combining them~\cite{zhang2025q}. 

Once the node has joined the network, it can send, and receive data. The data link layer frame for each transmission contains the MAC address of the destination. The size of this address is typically 64 bits or 8 bytes.         

The central idea that we employ is the possibility of reducing the number of bits required to identify the receiver (MAC address) by using ID codes. 

\subsection{ID codes for beacon messages}
Beacon messages enable neighbor discovery and maintenance in ad hoc networks by periodically broadcasting a node's identity and state to nearby devices~\cite{al2023afb}. When a node receives a beacon, it creates or updates an entry for that neighbor in its local neighbor table to track the topology. Maintenance relies on timeout mechanisms: if a node fails to receive a beacon from a specific neighbor within a predefined validity period, it assumes the neighbor has moved out of range and removes the entry from its table. Because ad hoc networks experience frequent topology changes, protocols like AODV and OLSR use these periodic hello messages to maintain up-to-date local connections and track neighbor relationships~\cite{al2023afb}.

We propose to use ID codes for the beacon messages for connectivity maintenance that are transmitted to the neighbors after the initial discovery. These messages are sent by the node with the ID code for its MAC address periodically. To verify the continued presence of a node within its neighbor list, a node monitors for periodic beacons containing the target node's ID.

Let us consider a practical example and consider the $6$-byte-long MAC address in the IEEE $802.11$ MAC frame~\cite{crow1997ieee}. With an ID code, suppose that the address length could be reduced to $1$ byte~\eqref{eq:doubleexpo}. Suppose that a node $A$ does not move from the neighborhood of node $B$ for $60$ minutes and is already in its neighbour list. Suppose that a beacon is sent approximately every 1000 ms~\cite{tan2005limiting}. Therefore the number of beacons sent over this one hour duration is approximately $\num{3.6e4}$. During this time, the strategy of using MAC address generates an order of magnitude more bytes than our strategy as shown in~\autoref{tab:sizecomp}. 
\begin{table}
\caption{Comparison when using complete MAC address and ID code in the beacon}\label{tab:sizecomp}
\centering
\begin{tabular}{||c c c||}  
 \hline
   & Using MAC & Using ID Code for MAC \\
 \hline
 Length  & 6 Bytes  & 1 Byte  \\
 Beacon Interval  & 1000ms  & 1000ms \\
Total Bytes in one hour & $2.16e4$ & $3.6e3$\\ 
\hline

\end{tabular}
\end{table}

This significant reduction in bits would be beneficial in energy limited systems. Since wireless transmission costs are much greater than computational costs~\cite{raghunathan2002energy}, this would be a highly valuable addition to the system.

\subsection{ID codes for short address in ad hoc networks}
We can also use the ID to shorten the MAC when sending data between two nodes that have already established connection. The sender encodes the receiver's MAC address and then uses this as the short address in the data link layer. The receiver checks if the ID is for itself, processes the frame, and otherwise discards it. Short addresses are used in practice mainly in low-power wireless sensor and IoT networks. IEEE 802.15.4 defines 16-bit short addresses in addition to 64-bit extended IDs. Short addresses are widely employed in IEEE 802.15.4-based wireless sensor networks and IoT systems, such as Zigbee-like mesh topologies, smart-city wireless sensor networks, and hierarchical IoT gateway architectures. Their use helps reduce header overhead and lower energy consumption, which is critical in battery-powered devices~\cite{kim2022cross}. Beyond basic addressing, these compact identifiers are integrated into specialized MAC protocols~\cite{kim2022cross}, utilized in compressed addressing schemes for high-density sensor deployments~\cite{schurgers2002distributed}, and serve as locally assigned MAC addresses~\cite{fan2018address} in Wi-Fi and ad hoc networks, where flexible, non-global identifiers offer practical advantages in dynamic or resource-constrained environments.

The short address in ZigBee is assigned by the PAN coordinator~\cite{ergen2004zigbee}. A coordinator-less network architecture offers significant advantages in rapidly changing network environments by eliminating dependence on a centralized PAN coordinator. This reduces the risk of a single point of failure and minimizes the control overhead associated with maintaining coordinator-node relationships when nodes frequently join, leave, or change position. Our proposed idea can be implemented in a coordinator-less network architecture where the devices themselves can generate the short address using the MAC address and ID coding.

\subsection{ID codes for path discovery}
When we consider a more general situation compared to the ring structure of ~\autoref{sec:packetrouting}, when a node has multiple neighbors and aims to forward a packet to only one path, routing algorithms are essential for ensuring efficient communication. There are two main types of routing protocols: table-driven routing protocols and on-demand routing protocols~\cite{royer1999review}. We look at the latter. On-demand routing in ad hoc networks operates by creating routes only when a source node requires communication with a destination. This approach is efficient in dynamic environments where nodes frequently move, join, or leave the network, and connections can be unreliable. When a node needs to send data, the on-demand routing algorithm initiates a route discovery process. This typically involves flooding the network with route requests until a path to the destination is found. Once a route is established, it is temporarily stored to facilitate subsequent communications, but only for as long as the route remains valid. 

We propose to use ID codes for path discovery. The node uses the ID of the destination address and sends the Route Request (RREQ). Upon receiving the RREQ, the destination node generates a Route Reply (RREP) message. This message is sent back to the source node, typically following the reverse path that the RREQ took to reach the destination. Note that we use MAC instead of IP for routing in the model so as to eliminate the overhead of IP headers, ARP broadcasts, and Network Layer processing, making it faster and more efficient.

Let's consider an example to illustrate this concept. Suppose we have a network path with 5 hops between the source and destination. In this scenario, instead of using the full 64-bit MAC address for destination, the transmitter uses only 16-bit ID. This represents a 75\% reduction in the destination address field and a similar reduction if applied to the source address field as well. This significant decrease in the header size can greatly improve network efficiency.
In essence, by using a ID instead of the full MAC address, we can substantially reduce the amount of data required to identify the source and destination in each packet. This can lead to better energy efficiency and more efficient network communication, as less data needs to be transmitted and processed at each hop.

\subsection{Common Randomness between the nodes}
As noted in~\autoref{sec:prereq}, the double-exponential advantage of ID is only possible with randomized ID. One method for implementing these ID codes is when the sender and receiver share common randomness~\cite{von2023beyond,gunlu2021code,derebeyouglu2020performance}. One way of sharing common randomness is to have random number generators (RNGs) in the receiver and transmitter initialized by the same seed~\cite{perotti2024identification}. This allows the receiver to produce the same pseudo-random integer as the transmitter’s RNG at any time instant. We propose to have this seed distributed confidentially between pairs of senders and receivers that want to communicate with one another. This also ensures that other nodes cannot decode the destination of a packet not intended for itself. For the smart warehouse in~\autoref{sec:packetrouting} for example, the initial seed can be shared between pairs of senders and receivers when they are deployed in the warehouse. This seed is then used to generate the ID code for the destination and the receiver also uses the same method to verify the received packet. This is similar to the procedure followed in~\cite{perotti2024identification}.

%% file: sec_challenges.tex
\section{Challenges and future work}\label{sec:concl}
Identification (ID) codes in post-Shannon communication promise huge gains in rate and traffic reduction, but their practical implementation faces several notable challenges.
Code construction complexity in ID is often larger than Shannon codes. Capacity-achieving ID codes often are mathematically and algorithmically more complex than standard Shannon-style transmission codes.
Computational cost is another disadvantage. The required pre-processing at the encoder and post-processing (verification) at the decoder can be significantly more demanding than conventional decoding, making high-throughput or low-latency implementations difficult.

ID codes also involve two types of errors (missed identification and false acceptance), and keeping both probabilities small in finite blocklengths is nontrivial, especially under realistic channel conditions.
While theory guarantees double-exponential identification capacity asymptotically, demonstrating clear practical gains at realistic blocklengths and SNRs remains a challenge and is an active area of research.

Our work focuses on introducing the use case of utilizing ID codes in ad hoc networks. Using ID for routing is not a replacement for the existing methods but a useful enhancement which allows for energy efficiency when used at the specific scenarios mentioned in our work. We find that a substantial header size reduction is possible when ID is used. This and the consequent reduction of wireless transmission power is particularly significant in energy limited systems such as wireless sensor networks. A complete simulation of the setup with the underlying changes in the protocols are part of the future work.